\documentclass[twocolumn,tighten,times]{aastex62}
\usepackage{graphicx}

\newcommand{\src}{G330.2+1.0}
\newcommand{\cha}{{\sl Chandra}}
\catcode`\@=11
\newcommand{\gapprox}{\mathrel{\mathpalette\@versim>}}
\newcommand{\lapprox}{\mathrel{\mathpalette\@versim<}}
\newcommand{\propapprox}{\mathrel{\mathpalette\@versim\propto}}
\newcommand{\@versim}[2]
  {\lower3.1truept\vbox{\baselineskip0pt\lineskip0.5truept
\ialign{$\m@th#1\hfil##\hfil$\crcr#2\crcr\sim\crcr}}}
\catcode`\@=12

\received{2018 August 25}
\revised{2018 October 23}
\accepted{2019 October 29}
\journalinfo{The Astrophysical Journal Letters, 2018, 868, L21}

\defcitealias{williams18}{W18}

\defcitealias{reynolds18}{R18}

\shorttitle{EXPANSION OF G330.2+1.0}

\begin{document}

\title{Expansion and Age of the X-ray Synchrotron-dominated Supernova
  Remnant G330.2+1.0}


\author{Kazimierz J. Borkowski}
\affiliation{Department of Physics, North Carolina State University, 
Raleigh, NC 27695-8202, USA}

\author[0000-0002-5365-5444]{Stephen P. Reynolds}
\affiliation{Department of Physics, North Carolina State University, 
Raleigh, NC 27695-8202, USA}

\author[0000-0003-2063-381X]{Brian J. Williams}
\affiliation{NASA/Goddard Space Flight Center, Greenbelt, MD 20771, USA}

\author[0000-0003-3850-2041]{Robert Petre}
\affiliation{NASA/Goddard Space Flight Center, Greenbelt, MD 20771, USA}

\begin{abstract}

  We report new {\sl Chandra} observations of one of the few Galactic
  supernova remnants whose X-ray spectrum is dominated by nonthermal
  synchrotron radiation, \src.  We find that between 2006 and 2017,
  some parts of the shell have expanded by about 1\%, giving a
  free-expansion (undecelerated) age of about 1000 yr, and
  implying shock velocities there of 9000 km s$^{-1}$ for a distance
  of 5 kpc.  Somewhat slower expansion is seen elsewhere around the
  remnant periphery, in particular in compact knots.  Because some
  deceleration must have taken place, we infer that \src\ is less than
  about 1000 yr old.  Thus, \src\ is one of only four Galactic
  core-collapse remnants of the last millennium.  The large size, low
  brightness, and young age require a very low ambient density,
  suggesting expansion in a stellar-wind bubble.  We suggest that in
  the east, where some thermal emission is seen and expansion
  velocities are much slower, the shock has reached the edge of the
  cavity.  The high shock velocities can easily accelerate
  relativistic electrons to X-ray-emitting energies.  A few small
  regions show highly significant brightness changes by $10\%$ --
  $20\%$, both brightening and fading, a phenomenon previously
  observed in only two supernova remnants, indicating strong and/or
  turbulent magnetic fields.

\end{abstract}


\section{Introduction}
\label{intro}

The fast shocks in young supernova remnants (SNRs) accelerate
particles to high enough energies to emit synchrotron radiation from
radio to X-ray wavelengths.  Studying such remnants can provide 
information on the detailed physics of the acceleration process.  In a
few remnants, synchrotron X-rays are not only present, but dominate
the spectrum (see \citealt{reynolds08} for a review).  While several
are quite well known, one lesser-known object has received recent 
attention, with a dominant synchrotron spectrum, thermal emission
in a few locations, and a compact central object (CCO), presumably a
neutron star: G330.2+1.0 \citep{park06,park09,torii06}.

\begin{figure}
   \includegraphics[width=3.5truein]{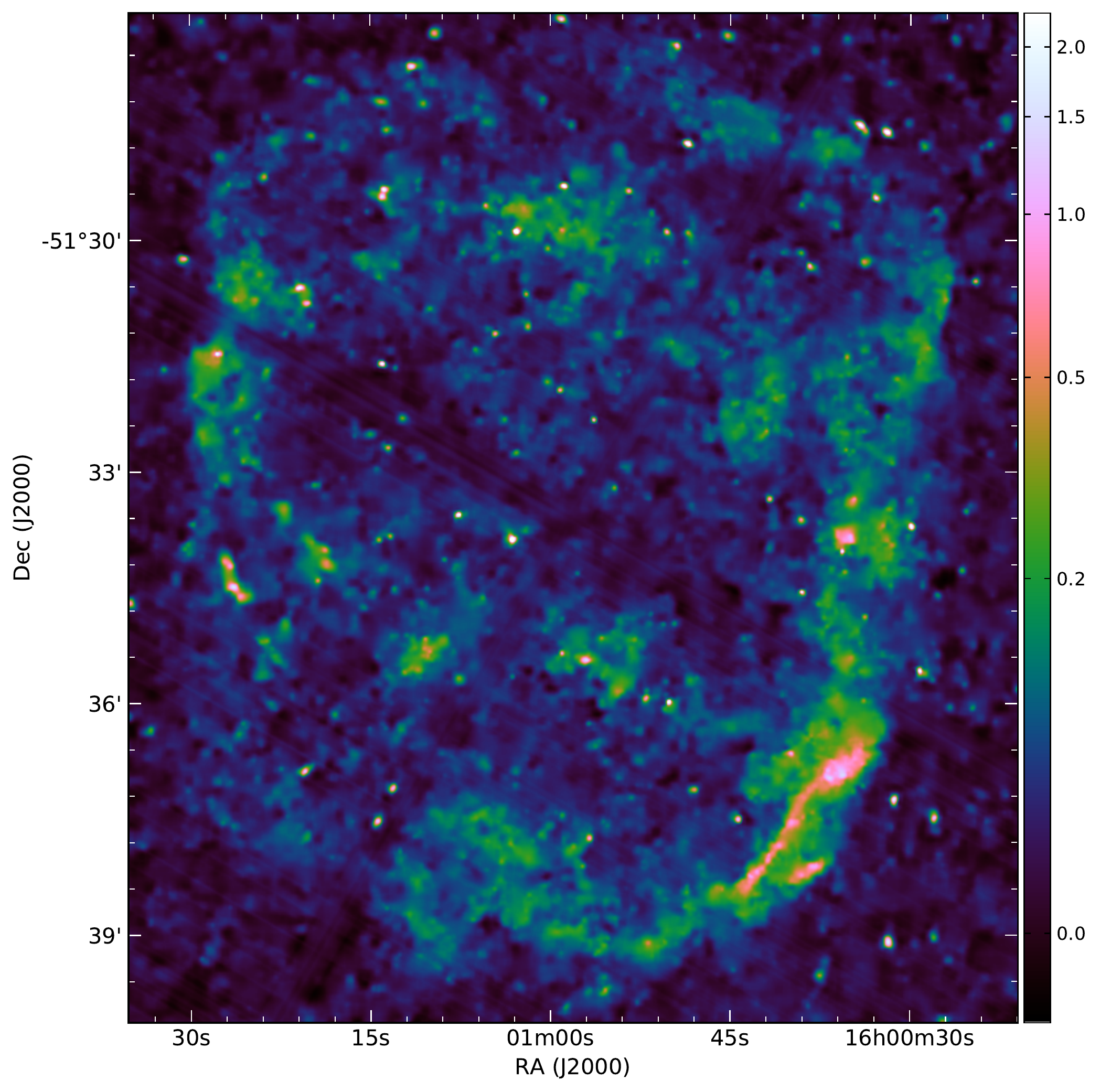}
  \caption{Smoothed background-subtracted {\sl Chandra} image of \src\ 
    (0.6--7.1 keV). The blast wave extends
    farthest away from the centrally located CCO toward the northwest. The
    bright southwest rim is inward facing. Compact emission knots are seen to
    the east and west of the CCO. The scale is in counts per
    $0 \farcs 769 \times 0 \farcs 769$ pixel. Intensities are shown with
    the cubehelix color scheme of \citet{green11}. The CCO is the bright
    point source near the center.}
\label{g330ch}
\end{figure}

\begin{figure}
   \includegraphics[width=3.5truein]{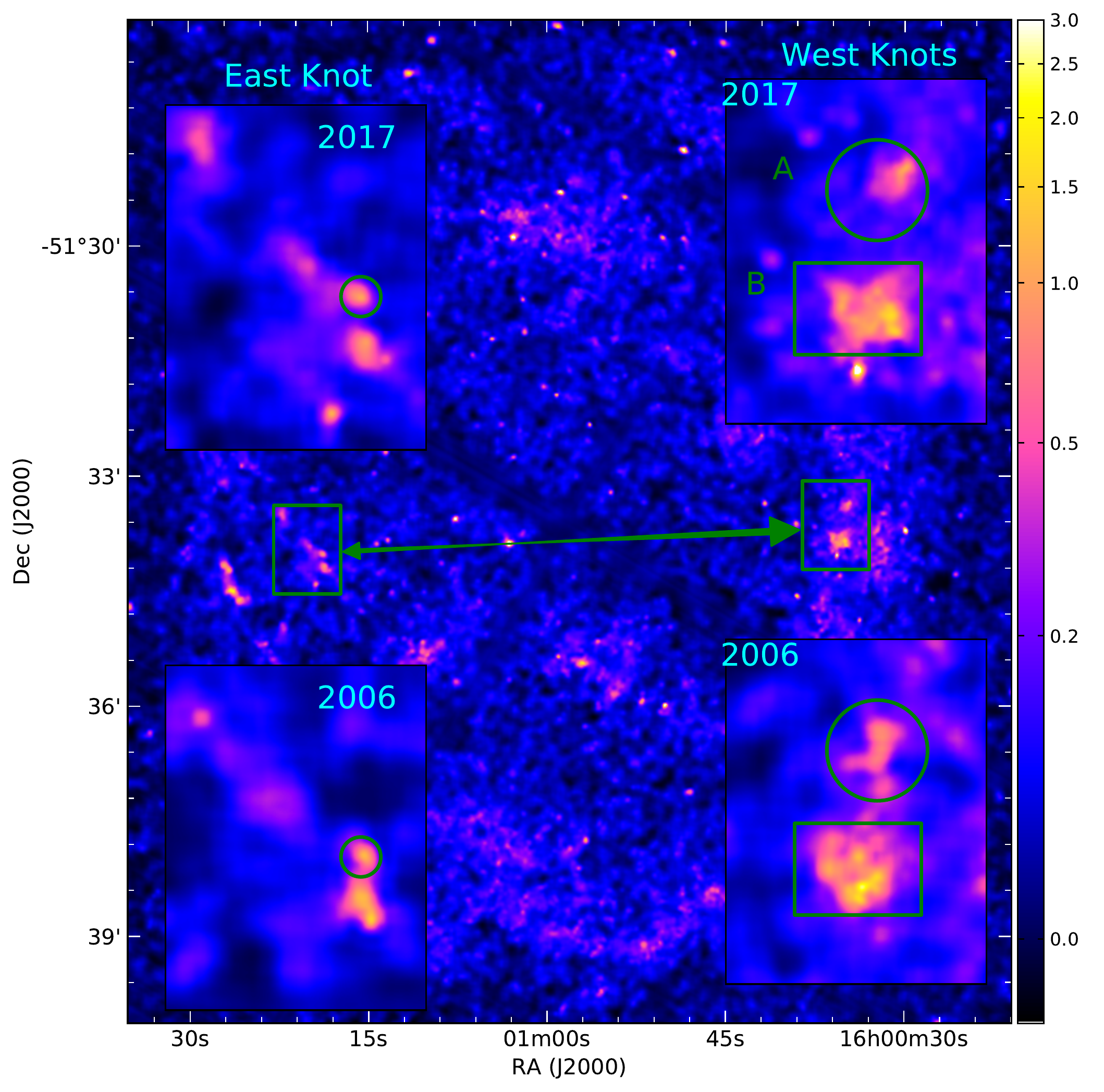}
   \caption{Zoom-ins of two regions (in green boxes on opposite sides of
     the CCO) containing 
     compact emission knots. Motions of two knots in the west
     (West Knots A and B), and at least one knot in the east (East Knot) are
     apparent between 2006 and 2017 (bottom and top insets, respectively).
     The scale is in counts per $0 \farcs 769 \times 0 \farcs 769$ pixel.}
\label{g330knotsfig}
\end{figure}

\begin{figure}
   \includegraphics[width=3.5truein]{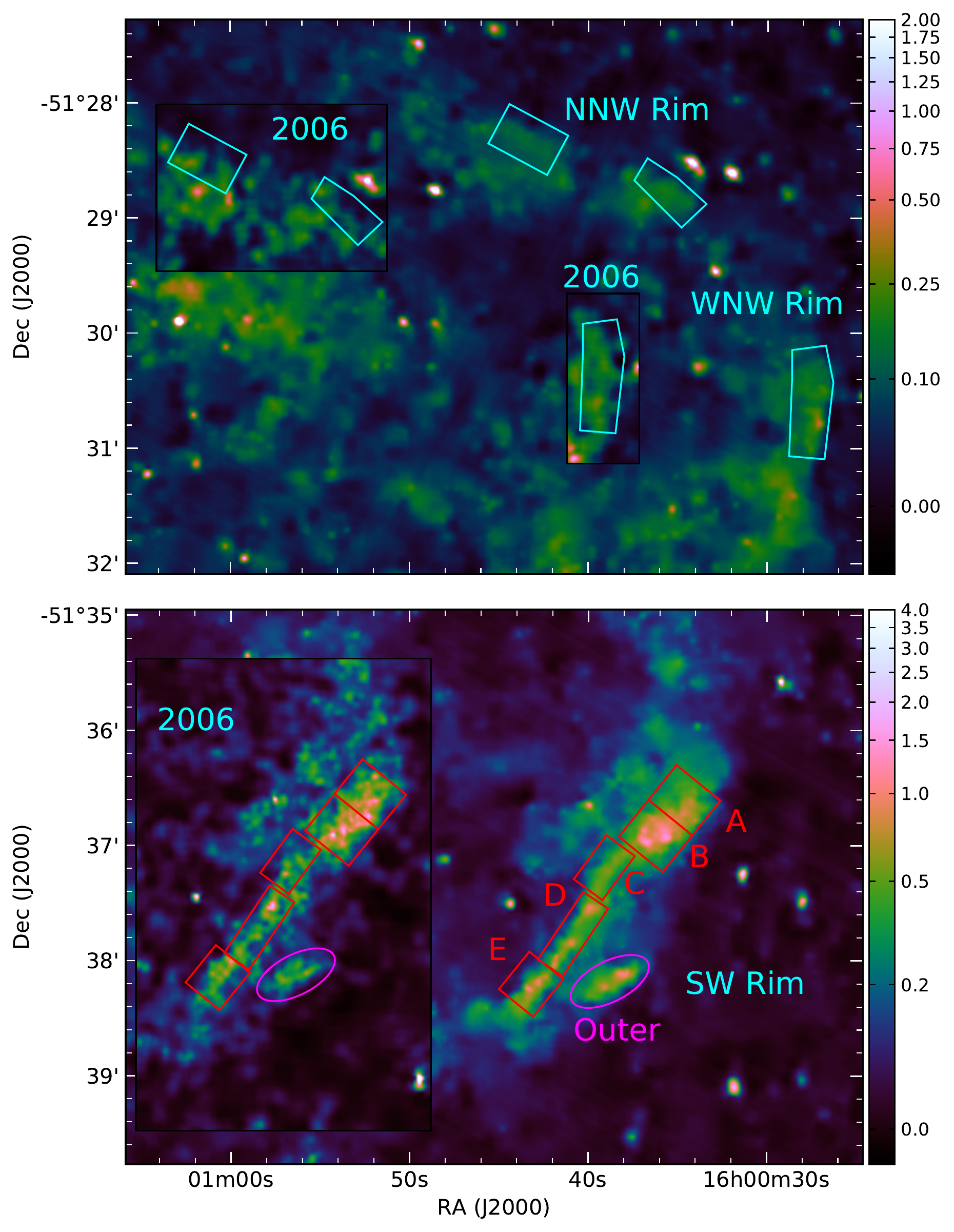}
   \caption{Zoom-ins of northwest (top) and southwest (bottom) rims of \src,
     with regions 
     used for expansion measurements of the blast wave overlaid. Insets show
     these regions in 2006. The scale is in the same units as in
     Figure~\ref{g330ch} (but insets are not to this scale).
  }
\label{g330blastwavefig}
\end{figure}

At radio wavelengths, \src\ has a distorted shell morphology with a
much brighter, confused region to the east \citep{caswell83}.
In X-rays
(Figure~\ref{g330ch}), the shell is clearer, coinciding fairly well
with much of the faint radio shell.

\src\ was observed with \cha\ \citep{park09} for 50 ks in 2006, and
with XMM-{\sl Newton} in 2008 \citep[][with an effective exposure 33
  ks]{park09} and 2015 \citep[][hereafter \citetalias{williams18},
  with effective exposure of about 90 ks]{williams18}.  Line emission
is virtually absent except in a region in the east (E) near the radio
maximum.  For the rest of the remnant, the X-ray spectrum between 1
and 7 keV can be well described by a single power law of photon index
$\Gamma \sim 2.3$ (\citealt{park09}; \citetalias{williams18}),
with little variation across the remnant, including not only the
clear shell, but emission from the faint interior
(\citetalias{williams18}).  These values are typical for power-law
components seen in other SNRs with nonthermal X-ray emission
\citep[e.g., RX J1713.7$-$3946;][]{katsuda15}.  However, the spectrum
could also be well described with a simple model of synchrotron
emission from a power-law energy distribution of electrons with an
exponential cutoff at some energy $E_m$ \citep[model {\tt srcut} in
  XSPEC;][]{reynolds99}.  The photon frequency corresponding to $E_m$
was found to be around $2 \times 10^{17}$ Hz, which is quite a high value; only
the youngest Galactic SNR G1.9+0.3 has a higher one, of
about $5 \times 10^{17}$ Hz \citep{reynolds09}.

H I absorption observations give a minimum distance of 4.9 kpc
\citep{mcclure-griffiths01}, the distance to some H I emission regions
with which it might be interacting.  While a distance as large as 10
kpc cannot be ruled out, it would imply unlikely properties such as an
age of 4000 yr for a synchrotron-dominated SNR, and a $z$-distance of 200 pc
above the Galactic plane for a core-collapse SNR. 
So we take a distance of $5\, d_5$ kpc, at which the mean angular
radius of about $5'$ corresponds to a linear radius of $7.3\,d_5$ pc.
\citetalias{williams18} use the thermal fits to the east region to obtain an
emission measure and an estimate of the preshock density in that
region of about $n_0 \sim 1\,d_5^{-1/2}$ cm$^{-3}$, but the absence of
thermal emission elsewhere suggests that this value is higher than
typical for the remnant environment. \citetalias{williams18} find an
electron temperature $T_e = 0.46 (0.40,0.58)$ keV for a shock velocity
of order 650 km s$^{-1}$; again, this is unlikely to typify the
remnant as a whole.

\section{Observations and Expansion Measurement Methods}
\label{obssec}

{\sl Chandra} observed \src\ again in two 74.1 ks exposures with the Advanced
Charge-coupled Device (CCD) Imaging Spectrometer (ACIS) I array on 2017 May 2
and 5,
in Very Faint mode. We used CIAO version 4.9 and CALDB version 4.7.8
to reprocess these observations, align them using the
centrally located CCO, and finally merge them together. No particle
flares were found. The total exposure time is 148.3 ks. A smoothed
X-ray image, extracted from merged and smoothed {\sl Chandra}
datacubes, is shown in Figure~\ref{g330ch}.
The remnant's faintness forces us to select only those
emission features that are 
bright and sharp enough to measure expansion reliably (see
Figures~\ref{g330knotsfig} and~\ref{g330blastwavefig} for their
location within the remnant).

The first-epoch {\sl Chandra} observations of \src\ in 2006, also
using the ACIS-I array in the Very Faint mode, were reprocessed in the
same way as the second epoch observations.
The exposure time is much shorter (50 ks), so the
uncertainties in our expansion measurements are almost entirely due to
the poor signal-to-noise ratio (S/N) of the first-epoch observations.
The CCO was used for
inter-epoch alignment, but a point source reference frame was also 
examined,
in both cases using the same techniques as in our expansion measurements of
the 500 yr old pulsar wind nebula in the SNR Kes 75
\citep[][hereafter \citetalias{reynolds18}]{reynolds18}.
There are 18 sources with off-axis angle $\theta_{\rm OAA}<7'$,
encompassing our selected emission features, that are suitable
for the astrometric alignment. Our null hypothesis is that their displacements
between the two epochs, measured after alignment to the CCO and normalized by
$1 \sigma$ errors estimated as in \citetalias{reynolds18}, satisfy the
Rayleigh distribution that is expected if the alignment were
perfect. The Kolmogorov--Smirnov test fails to reject this 
hypothesis with $82\%$ confidence. The CCO and point source
reference frames are displaced by only 77 mas when measured using point
sources with $\theta_{\rm OAA} < 5'$ where our source
position measurements are most reliable. The estimated
\citepalias[as in][]{reynolds18} positional uncertainties for
the CCO are only slightly smaller, about 60 mas
(at $90\%$ confidence) for each pointing.
Therefore, we find no evidence for misalignment
of point sources or a discernible CCO motion. 

The time baseline between the 2006 and 2017 observations is 10.95 yr,
long enough to reliably measure expansion of most of the selected
emission features with a variation of the method used by
\citetalias{reynolds18}.
First, we extracted a datacube from the
merged 2017 data, with $1024^2$ image pixels and 16 spectral channels,
in the energy range from $0.6$ to $7.1$ keV. The spatial pixel size is
$0 \farcs 769 \times 0 \farcs 769$ (slightly more than $3/2$ of an
ACIS $0 \farcs 492$ pixel). We smoothed this datacube with the
multiscale partitioning method of \citet{krishnamurthy10}, and
then extracted smoothed images.
Three relatively high-surface-brightness
emission knots (Figure~\ref{g330knotsfig}), one in the east and two in
the west, are significantly more compact than more diffuse rims and
filaments in north-northwest, west-northwest, and southwest
(Figure~\ref{g330blastwavefig}). The East
Knot is the most compact (FWHM of $2 \farcs 9$). In order to avoid oversmoothing
this and the other two knots, the penalty parameter that
controls smoothing was set to $0.03$ in this case (the smoothed image is
shown in
Figure~\ref{g330knotsfig}) vs.~$0.06$ for the rest of G330.2+1.0 
(Figures~\ref{g330ch} and ~\ref{g330blastwavefig}). 
(Oversmoothing of the knots using the penalty parameter of $0.06$ 
simply degrades our measurements; e.g., the expansion variance for the East
Knot increases by over $60\%$ without any significant change in the measured
expansion.) After particle background subtraction and
normalization by a monochromatic ($E=2.3$ keV) exposure map, these
smoothed images are used as models for the brightness distribution of
the selected emission features of interest.

The smoothed 2017 models are fit to the unsmoothed $0.6$--$7.1$ keV
image from 2006 using Markov chain Monte Carlo (MCMC) methods as
implemented in the PyMC software package \citep{patil10}. (But we show
parts of a smoothed 2006 image in Figures~\ref{g330knotsfig}
and~\ref{g330blastwavefig} instead of the unsmoothed image -- the
penalty parameter was set to $0.02$ in smoothing the 2006 datacube
from which this smoothed image was extracted.) Poisson statistics are
assumed. In our MCMC simulations, we allow for changes in the physical
image scale and in the surface-brightness scale factor $S$.  Expansion
is centered on the CCO. Spatial variations in the effective exposure
time are accounted for by the monochromatic ($E=2.3$ keV) exposure
map. The particle background is modeled as
described in \citet{bartalucci14}. 
A uniform prior is assumed for expansion, reflecting
our prior (lack of) knowledge about the age and dynamics of \src.

Approximately, the scale factor $S$ can be considered as a proxy for
the (scaled) photon flux within each of our regions.
We use a Gaussian prior for $S$ with mean of unity and variance
$\sigma_S^2$.  The lower bound to $\sigma_S$ is $(N_s+N_b)^{1/2}/N_s$
(where $N_s$ and $N_b$ are the total number of source and background
counts within each of our regions in 2017). We conservatively assume
the variance $\sigma_S^2$ to be twice as large, equal to
$2(N_s+N_b)/N_s^2$. But for three regions in the southwest that unexpectedly
showed evidence for apparent flux variations between the two epochs,
Southwest Rims A and B, and Southwest Outer Rim, we further relaxed the prior
for $S$ by setting $\sigma_S^{-2}$ to $30$. We list $\sigma_S$ estimated
in this way in the second column of Table~\ref{rates}.

For each region, our MCMC simulations involved about 15 chains, each
4000 iterations in length.
We list chain-averaged means for
expansion (and expansion rate) and $S$ in Table~\ref{rates}.  Approximate
radial distances measured from images, together with radial proper motions
and velocities, are also listed there. Two sets of credible 
intervals are provided, corresponding to $(16\%,84\%)$ and
$(2.5\%,97.5\%)$ quantiles from MCMC draws. Unless noted otherwise, we
refer to the first set (corresponding to $1\sigma$ errors) when
quoting errors on expansion, $S$, and on other derived quantities.

\section{Results and Discussion}

\begin{deluxetable*}{lccccccc}
\tablecolumns{8}
\tablecaption{Expansion between 2006 and 2017\label{rates}}
\tablehead{
\colhead{Region} & $\sigma_S$\tablenotemark{a} & $S$\tablenotemark{b} & Expansion & Expansion Rate & Distance & $\mu_r$\tablenotemark{c} & $v_5$\tablenotemark{d} \\
& & & (\%) & (\%~yr$^{-1}$) & (arcsec) & (arcsec yr$^{-1}$) & (km s$^{-1}$) }

\startdata
North-northwest Rim     & $0.11$ & $0.95$ & $0.98$ & $0.089$ & $395$ & $0.35$ & $8400$ \\
& & $(0.86,1.04)$ & $(0.64,1.31)$ & $(0.059,0.120)$ & & $(0.23,0.47)$ & $(5500,11,\! 000)$ \\
& & $(0.78,1.12)$ & $(0.20,1.60)$ & $(0.018,0.146)$ & & $(0.07,0.58)$ & $(1700,14,\! 000)$ \\
West-northwest Rim     & $0.10$ &  $1.01$ & $1.07$ & $0.098$ & $390$ & $0.38$ & $9100$ \\
& & $(0.93,1.10)$ & $(0.77,1.37)$ & $(0.070,0.125)$ & & $(0.27,0.49)$ & $(6500,12,\! 000)$\\
& & $(0.85,1.17)$ & $(0.44,1.63)$ & $(0.041,0.149)$ & & $(0.16,0.58)$ & $(3700,14,\! 000)$ \\
West Knot A & $0.15$ &  $1.13$ & $1.36$ & $0.125$ & $265$ & $0.33$ & $7800$ \\
& & $(1.02,1.24)$ & $(0.94,1.71)$ & $(0.086,0.157)$ & & $(0.23,0.41)$ & $(5400,9800)$ \\
& & $(0.92,1.34)$ & $(0.60,2.01)$ & $(0.055,0.183)$ & & $(0.15,0.49)$ & $(3400,12,\! 000)$ \\
West Knot B & $0.08$ &  $1.08$ & $0.89$ & $0.082$ & $260$ & $0.21$ & $5000$ \\
& & $(1.02,1.15)$ & $(0.77,1.01)$ & $(0.071,0.092)$ & & $(0.18,0.24)$ & $(4400,5700)$ \\
& & $(0.96,1.20)$ & $(0.66,1.14)$ & $(0.060,0.104)$ & & $(0.16,0.27)$ & $(3700,6400)$ \\
Southwest Rim A    & $0.18$ &  $1.23$ & $0.54$ & $0.049$ & $315$ & $0.16$ & $3700$ \\
& & $(1.16,1.30)$ & $(0.34,0.74)$ & $(0.031,0.068)$ & & $(0.10,0.21)$ & $(2300,5100)$ \\
& & $(1.09,1.37)$ & $(0.14,0.91)$ & $(0.013,0.083)$ & & $(0.04,0.26)$ & $(1000,6200)$ \\
Southwest Rim B    & $0.18$ &  $1.12$ & $0.00$ & $0.000$ & $310$ & $0.00$ & $0$ \\
& & $(1.06,1.19)$ & $(-0.18,0.17)$ & $(-0.016,0.016)$ & & $(-0.05,0.05)$ & $(-1200,1200)$ \\
& & $(1.00,1.25)$ & $(-0.38,0.34)$ & $(-0.035,0.031)$ & & $(-0.11,0.10)$ & $(-2500,2300)$ \\
Southwest Rim C    & $0.09$ &  $0.97$ & $-0.33$ & $-0.030$ & $300$ & $-0.09$ & $-2200$ \\
& & $(0.90,1.04)$ & $(-0.72,0.03)$ & $(-0.066,0.003)$ & & $(-0.20,0.01)$ & $(-4700,200)$ \\
& & $(0.84,1.11)$ & $(-1.11,0.30)$ & $(-0.101,0.027)$ & & $(-0.30,0.08)$ & $(-7200,1900)$ \\
Southwest Rim D    & $0.07$ &  $1.07$ & $0.38$ & $0.034$ & $315$ & $0.11$ & $2600$ \\
& & $(1.02,1.12)$ & $(0.22,0.53)$ & $(0.020,0.048)$ & & $(0.06,0.15)$ & $(1500,3600)$ \\
& & $(0.96,1.18)$ & $(0.05,0.65)$ & $(0.005,0.060)$ & & $(0.02,0.19)$ & $(400,4400)$ \\
Southwest Rim E    & $0.09$ &  $1.01$ & $0.71$ & $0.065$ & $320$ & $0.21$ & $4900$ \\
& & $(0.95,1.08)$ & $(0.54,0.89)$ & $(0.049,0.081)$ & & $(0.16,0.26)$ & $(3700,6200)$ \\
& & $(0.89,1.14)$ & $(0.29,1.04)$ & $(0.027,0.095)$ & & $(0.08,0.30)$ & $(2000,7200)$ \\
Southwest Outer Rim & $0.18$ &  $0.81$ & $0.24$ & $0.022$ & $345$ & $0.07$ & $1800$ \\
& & $(0.74,0.88)$ & $(0.07,0.41)$ & $(0.006,0.037)$ & & $(0.02,0.13)$ & $(500,3000)$ \\
& & $(0.67,0.96)$ & $(-0.11,0.54)$ & $(-0.010,0.050)$ & & $(-0.03,0.17)$ & $(-800,4100)$ \\
East Knot   & $0.25$ &  $1.14$ & $0.91$ & $0.083$ & $145$ & $0.12$ & $2900$ \\
& & $(0.96,1.33)$ & $(0.46,1.34)$ & $(0.042,0.122)$ & & $(0.06,0.18)$ & $(1500,4200)$ \\
& & $(0.79,1.51)$ & $(0.11,1.75)$ & $(0.010,0.160)$ & & $(0.01,0.23)$ & $(300,5500)$ \\
\enddata
\tablecomments{Credible intervals are two sets of $(16\%,84\%)$ and
  $(2.5\%,97.5\%)$ quantiles from MCMC draws.}
\tablenotetext{a}{Width of Gaussian prior for the surface brightness scaling
  factor $S$ (its mean was set to unity).}
\tablenotetext{b}{Surface brightness scaling factor.}
\tablenotetext{c}{Radial proper motion.}
\tablenotetext{d}{Expansion velocity is $v_5d_5$, with $d_5$ the distance in units of 5 kpc.}
\end{deluxetable*}

\src\ shows a highly irregular blast wave and a number
of compact emission knots to the east and west of the CCO
(Figure~\ref{g330ch}). We
highlight three such emission knots in Figure~\ref{g330knotsfig}, West
Knots A and B, and the East Knot. They have contrasting spectral
properties, with West Knots having hard spectra typical of emission in
the western part of \src, thus being most prominent in the hard ($2$--$7$ keV)
energy band in Figure 1 of \citetalias{williams18} (see their region $10$).
The East Knot is in the inner part of a region labeled as ``therm'' in this
Figure, within the most distinct knot of thermal emission there that is
very prominent in the $1.2$--$2$ keV band, so its
spectrum is presumably thermal. All three knots are moving away from the CCO
with comparable expansion rates: $0.125_{-0.039}^{+0.032}\%$ yr$^{-1}$ and
$0.082\% \pm 0.011\%$ yr$^{-1}$ for the West Knots A and B, and
$0.083_{-0.041}^{+0.039}\%$ yr$^{-1}$ for the East Knot
(Table~\ref{rates}). We identify the East Knot with a dense clump of
supernova (SN) ejecta because of its compactness, soft spectrum, and
high transverse motion of $2900_{-1400}^{+1400}d_5$ km s$^{-1}$. The
West Knots might also be associated with fast-moving ejecta clumps,
although their densities are probably much lower. Because their
velocities are much higher, $5000_{-700}^{+700}d_5$ km s$^{-1}$ for the West
Knot B and possibly even more for the West Knot A, nonthermal
emission is likely to be more efficiently produced, perhaps accounting
for the profound differences in X-ray spectral properties between the
East and West Knots.

The East and West Knots are located on opposite sides of the CCO. 
This allows us to relax our approximation that the CCO has not 
moved away from the SN explosion site, by allowing these knots to expand 
uniformly around a common expansion center that may not coincide with the CCO.
A 2D Gaussian prior with the
FWHM of $25\arcsec$ centered on the CCO was used for the expansion center.
Its width
was chosen to exclude unrealistically large ($>30\arcsec$) displacements 
produced by a hypothetical neutron star with age longer than 1500 yr that is
moving with the transverse velocity of
$500d_5$ km s$^{-1}$ or more.
We obtain an expansion
rate of $0.085_{-0.010}^{+0.011}\%$ yr$^{-1}$, virtually identical to
$0.083_{-0.009}^{+0.010}\%$ yr$^{-1}$ obtained with the expansion
centered on the CCO.
Allowing for the CCO motion results in only a modest loss in precision, and
a statistically insignificant increase in expansion. 

\begin{figure}
   \includegraphics[width=3.4truein]{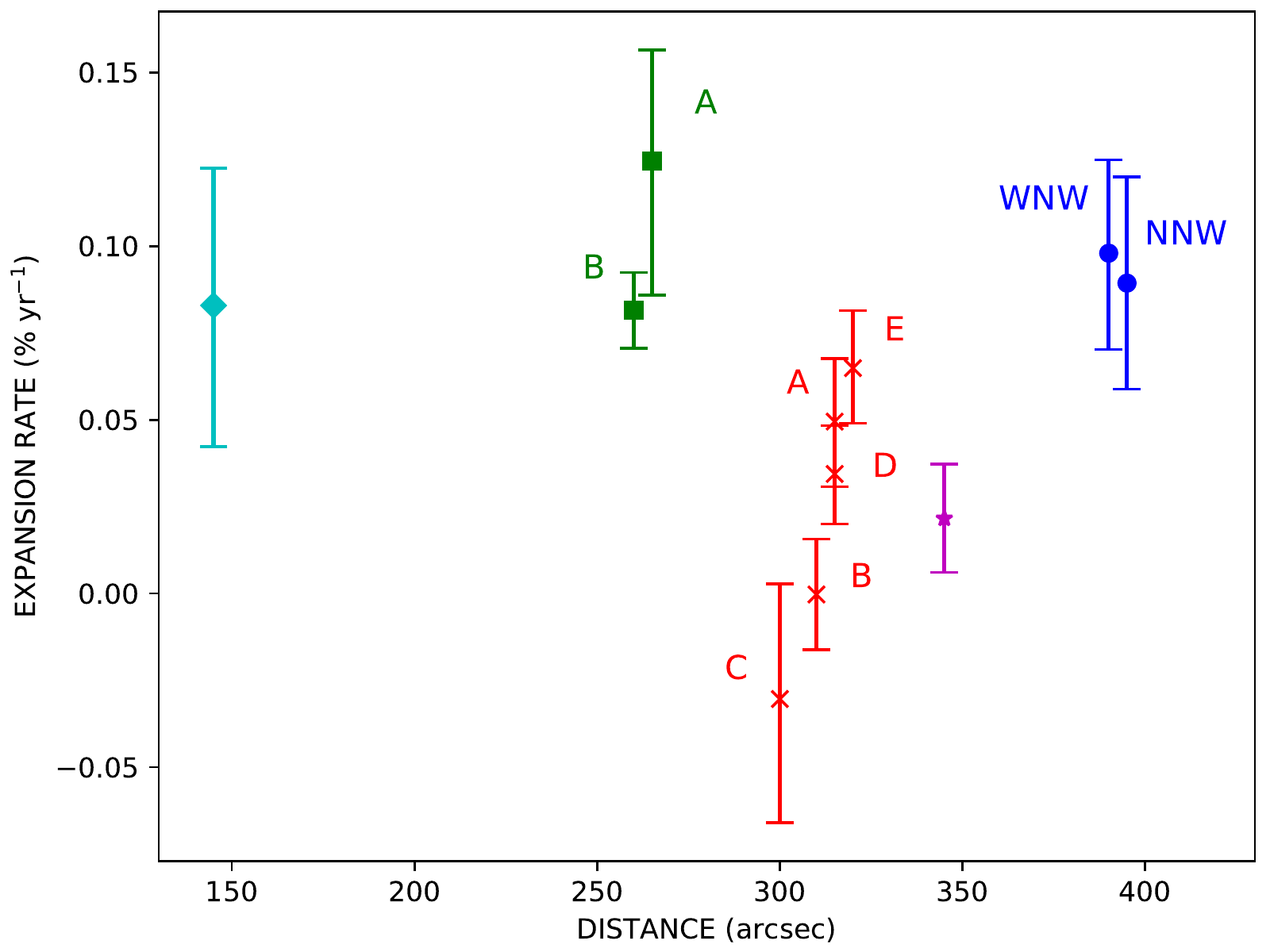}
   \caption{Expansion rate vs.~radial distance from the CCO. From left to
     right: East Knot (cyan diamond), West Knots (green squares), Southwest
     Rim (red $\times$'s), Southwest Outer Rim (magenta star) and
     West-northwest and North-northwest Rims (blue circles). A large range in
     expansion is apparent in the southwest.
  }
\label{g330expansion}
\end{figure}

The strongly nonspherical blast wave extends farthest away from the
CCO toward the northwest, with sharp rims visible in the north-northwest and
west-northwest (upper
panel in Figure~\ref{g330blastwavefig}). 
While a chip gap divides the north-northwest rim,
expansion was measured by combining the two apparently noncontiguous
rim sections. We find expansion rates of $0.089\% \pm
0.031\%$ yr$^{-1}$ and $0.098_{-0.028}^{-0.027}\%$ yr$^{-1}$ in north-northwest 
and west-northwest, virtually identical within errors (the weighted expansion
rate is $0.094\%$ yr$^{-1}$), and also comparable to the expansion
found for the ejecta knots in the east and west
(Figure~\ref{g330expansion}). But the blast wave velocities of
$9100_{-2600}^{+2500}d_5$ km s$^{-1}$ in west-northwest and 
$8400_{-2900}^{+2900}d_5$ km s$^{-1}$ in north-northwest 
are much larger than velocities of the
compact knots.

The bright rim in the southwest of \src\ has a complex morphology, with an
inward-facing rim and a detached outer section that is perhaps not
physically related.
Figure~\ref{g330blastwavefig} (lower panel) shows five regions spanning
the entire bright inward-facing rim and an additional outer region
where we measured radial expansion. We found no measurable motions
near the apex of the Southwest Rim closest to the CCO, but a significant
expansion is detected on either side of it (Table~\ref{rates} and
Figure~\ref{g330expansion}).  The fastest expansion is found at the northwest 
and southeast tips of the Sothwest Rim (labeled A and E in
Figure~\ref{g330blastwavefig}): $0.049_{-0.019}^{+0.018}\%$ yr$^{-1}$
and $0.065_{-0.016}^{+0.017}\%$ yr$^{-1}$, respectively. This is still
less than in the northwest or for the ejecta knots, but the radial velocity
of $4900_{-1200}^{+1300}d_5$ km s$^{-1}$ at its southeast tip is still impressive.

The scale factor $S$ is significantly above unity in the Southwest Rim northwest
of its apex, and below unity for the Southwest Outer Rim (Table~\ref{rates}).
The former includes the brightest part of \src, the Southwest Rim B, that faded
by $12\%$. Even more pronounced ($23\%$) fading is found for the adjacent
Southwest Rim A, while the Southwest Outer Rim brightened by a comparable
($19\%$) amount. Such large
flux variations cannot be accounted for by either statistical errors
(see Table~\ref{rates}) or by systematic calibration errors that can reach only 
about $3\%$\footnote{\scriptsize http://cxc.harvard.edu/cal/summary/Calibration\_Status\_Report.html\#ACIS\_EA} for a highly absorbed source such as \src.
X-ray
flux variations of any kind are rare in shell SNRs.
Small-scale variations have been reported in
the synchrotron-dominated shell remnant
G347.3$-$0.5 \citep[RX J1713.7$-$3946;][]{uchiyama07} while SN 1006,
also synchrotron-dominated, shows no such variations
\citep{katsuda10}.  Such variations seem to require magnetic
fields strongly amplified in the shock waves \citep{uchiyama07}, but
turbulent fields need not be as large on average \citep{bykov08}.

The very fast blast wave propagating through a low-density medium such as
encountered in the northwest must have hit a dense obstacle in the
southwest some 
time ago. The blast wave first arrived near the apex of the Southwest Rim,
slowed down there, then started to wrap around this
obstacle. Qualitatively, the observed expansion pattern and the
overall geometry of the Southwest Rim are consistent with this
scenario. Rapid brightness variations might also be
expected for a relatively recent interaction.

Our expansion rates in the north and west of order 0.1\% yr$^{-1}$
give an undecelerated age of 1000 yr (800 yr for the fastest
knot, West Knot A).  As some deceleration has almost certainly
occurred, these are upper limits.  Evidently \src\ is one of the
youngest SNRs in the Galaxy, probably younger even than the
synchrotron-X-ray-dominated Type Ia remnant SN 1006.  Along with Cas
A, Kes 75 \citepalias{reynolds18}, and the Crab, \src\ is one of the four
youngest Galactic core-collapse remnants known.  Even with the
uncertain distance, the shock speeds that we measure are among the fastest
ever measured for an SNR.  Only G1.9+0.3, with speeds of
order 14,000 km s$^{-1}$ \citep{carlton11}, has significantly faster
shocks.

The large average size ($7.3 d_5$ pc), the young age of less than 1000
yr, very high shock speeds up to 9000 km s$^{-1}$, and very low flux
imply a very low mean density of ambient material that is typical of
wind-blown bubbles around massive stars.  However, the presence of
thermal emission in the east quadrant indicates a considerably higher
density there \citepalias{williams18}.  We suggest that \src\ is
expanding in an asymmetric cavity, likely produced by
pre-supernova mass loss, but that in the east, the blast wave has
reached the cavity wall and is expanding into much denser material.
Then \src\ is the first core-collapse SNR for which the interaction
with a cavity wall has begun but is not complete, similar to the
(probable) Type Ia remnant RCW 86 \citep{williams11}. A number of
small-scale emission features are found in the east (Figure~\ref{g330ch}),
suggesting a considerable amount of clumpiness within or near the cavity wall.

The high shock velocities that we find can easily account for the high
rolloff frequencies of the synchrotron emission.  If electron
acceleration is limited by radiative losses, that frequency is
independent of magnetic field: $\nu_{\rm rolloff} = 5 \times 10^{16}
u_8^2 (\eta R_J(\theta_{Bn}))^{-1}$ Hz, where $u_8$ is the shock
velocity in units of $10^8$ cm s$^{-1}$, $\eta$ is the scattering mean
free path of a relativistic electron in units of its gyroradius (the
``gyrofactor''), and $R_J(\theta_{Bn})$ is a factor accounting for
obliquity-dependence of shock acceleration \citep{reynolds98}.  For a
strong, turbulent shock we expect $\eta R_J \sim 1$.  The observed
values of rolloff frequency, about $3 \times 10^{17}$ Hz, then require
only $u_8 \gapprox 3,$ lower than most of the velocities that we measure.
The shock velocities are so high that the age-limited 
maximum electron energy, which varies as $u_8^4$, is much higher than 
the loss-limited energy and hence not relevant.

\citetalias{williams18} find no appreciable variation in rolloff
frequencies among 14 positions around the rim, with uncertainties of
order (60 -- 80)\%, suggesting that unlike SN 1006, where systematic
variations with azimuth of an order of magnitude are seen
\citep{miceli09} and attributed to systematic variations in shock
obliquity, the upstream magnetic-field direction outside \src\ is far
from uniform.

\section{Conclusions}

We have measured the expansion of the X-ray-synchrotron-dominated SNR
\src\ between 2006 and 2017.  We find expansion rates of up to 0.12\%
yr$^{-1}$, for an undecelerated age of less than 1000 yr. Because 
some deceleration is certain to have occurred, the true age is even
less.  \src\ is thus among the youngest core-collapse SNRs in the
Galaxy, and has shock velocities faster than all known SNRs except the
youngest Galactic SNR G1.9+0.3, thought to be Type Ia, whose X-rays
are also dominated by synchrotron emission.  The shock velocities in
\src\ range up to 9000 km s$^{-1}$, in regions somewhat farther from
the center than the average radius, where we presume the central
neutron star to be the expansion center.  Some distinct blobs have
somewhat slower expansion rates, corresponding to speeds of about 3000
-- 5000 km s$^{-1}$, and softer spectra, suggestive of clumps of
ejecta.  Complex motions are seen in the southwest, where some regions have
varied in brightness by 10\% -- 20\%. The presence of thermal emission
in one region in the east suggests interaction with much denser
material there.  We propose that \src\ is expanding into a
stellar-wind cavity and has reached the cavity wall in the east some
time ago.  A more recent impact with the cavity wall might have also
occurred in the southwest in view of the strong blast wave deceleration and
the presence of significant brightness changes there.  Our age of
order 1000 yr or less may be significant for the cooling of the
neutron star, which is evidently the youngest CCO after the CCO
in Cas A.

\acknowledgments We acknowledge support by NASA through {\sl Chandra}
General Observer Program grant SAO GO7-18051X.

\vspace{5mm}
\facilities{CXO}

\end{document}